\documentclass[runningheads]{llncs}
\usepackage[utf8]{inputenc}
\usepackage{amsmath,amssymb}
\usepackage{booktabs}
\usepackage{subfig}
\usepackage{subcaption}
\usepackage{graphicx}
\usepackage[T1]{fontenc}
\usepackage{algorithm}
\usepackage{algpseudocode}

\usepackage{color}
\usepackage{bbding}

\begin{document}
%
\title{Lightweight Hypercomplex MRI Reconstruction: A Generalized Kronecker-Parameterized Approach}
\titlerunning{Lightweight Hypercomplex MRI Reconstruction}

%



\author{
Haosen Zhang\inst{1, *} \and
Jiahao Huang\inst{1, 3, 4, *, \dagger} \and
Yinzhe Wu\inst{1, 3, 4, *} \and
Congren Dai\inst{2} \and
Fanwen Wang\inst{1, 3, 4} \and
Zhenxuan Zhang\inst{1} \and
Guang Yang\inst{1,3,4,5, \dagger}
}

\authorrunning{Zhang et al.}

\renewcommand{\thefootnote}{\fnsymbol{footnote}}
\footnotetext[1]{These authors contributed equally to this work. $\dagger$ Co-corresponding author.}

\institute{
Bioengineering Department and Imperial-X, Imperial College London, London, UK \\
\and Department of Computing, Imperial College London, London, UK \\
\and National Heart and Lung Institute, Imperial College London, London, UK \\
\and Cardiovascular Research Centre, Royal Brompton Hospital, London, UK \\
\and School of Biomedical Engineering and Imaging Sciences, King’s College London, UK \\
\texttt{j.huang21@imperial.ac.uk, g.yang@imperial.ac.uk}
}

\maketitle              
\begin{abstract}
Magnetic Resonance Imaging (MRI) is crucial for clinical diagnostics but is hindered by prolonged scan times. Current deep learning models enhance MRI reconstruction but are often memory-intensive and unsuitable for resource-limited systems. This paper introduces a lightweight MRI reconstruction model leveraging Kronecker-Parameterized Hypercomplex Neural Networks to achieve high performance with reduced parameters. By integrating Kronecker-based modules, including Kronecker MLP, Kronecker Window Attention, and Kronecker Convolution, the proposed model efficiently extracts spatial features while preserving representational power. We introduce Kronecker U-Net and Kronecker SwinMR, which maintain high reconstruction quality with approximately 50\% fewer parameters compared to existing models. Experimental evaluation on the FastMRI dataset demonstrates competitive PSNR, SSIM, and LPIPS metrics, even at high acceleration factors (8$\times$ and 16$\times$), with no significant performance drop. Additionally, Kronecker variants exhibit superior generalization and reduced overfitting on limited datasets, facilitating efficient MRI reconstruction on hardware-constrained systems. This approach sets a new benchmark for parameter-efficient medical imaging models.


\end{abstract}

\section{Introduction}
\begin{figure}[htbp]
    \centering
    \includegraphics[width=1\textwidth]{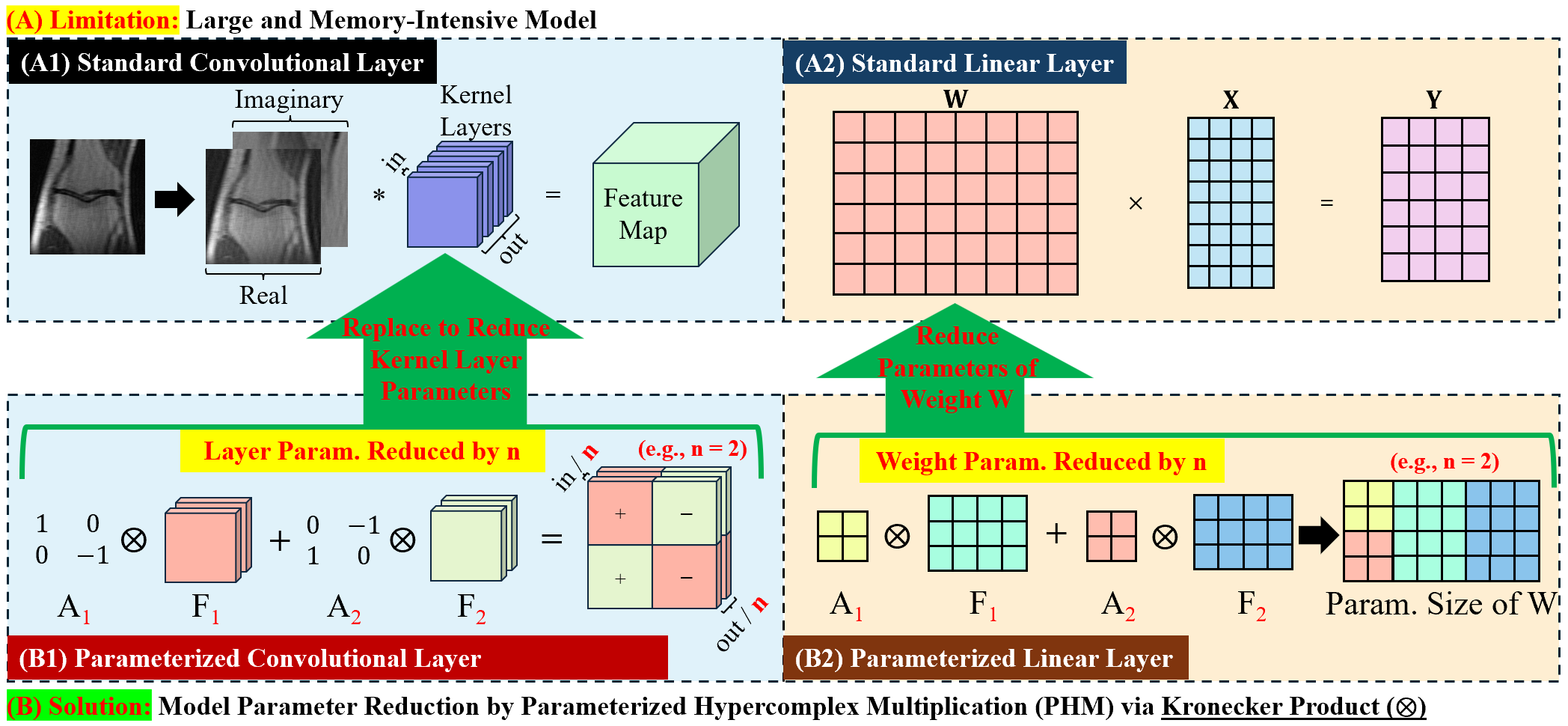}
    \caption{(A) Many state-of-the-art deep learning MRI reconstruction models are large and memory-intensive. To reduced the parameters of (A1) Convolutional Layers and (A2) Linear Layers, (B) they were redevised through Parameterized Hypercomplex Multiplication (PHM) through Kronecker Product. A hypercomplex dimension of n (n=2 here) in PHM would reduce model parameters by the same factor in Parameterized (B1) Convolutional Layer and (B2) Linear Layer. }
    \label{fig:framework}
\end{figure}

Magnetic Resonance Imaging (MRI) is a non-invasive imaging technique essential for clinical diagnosis but is limited by long scan times. Parallel imaging and compressed sensing accelerate scans by reconstructing undersampled k-space data. Recently, deep learning has enhanced MRI reconstruction by inferring missing k-space, achieving higher image fidelity ~\cite{Hammernik2023physics_driven,huang2025data_physiscs}.

Despite significant advancements, state-of-the-art deep learning models for MRI reconstruction are challenging to deploy in clinical settings due to their large size and high memory requirements. Even efficient models like SwinMR~\cite{huang2022swintransformerfastmri,9607618} have around 11 million parameters, demanding substantial GPU memory and computational power. This makes real-time imaging on resource-constrained MRI scanners impractical~\cite{Xu_LBUNet_MICCAI2024}. Consequently, there is growing interest in lightweight neural networks that provide fast inference within the hardware limitations of clinical scanners.

Compact Convolutional Neural Networks (CNNs) and efficient learning strategies have been developed to address this challenge. EfficientNet~\cite{pmlr-v97-tan19a,pmlr-v139-tan21a} achieves state-of-the-art accuracy with significantly fewer parameters by scaling network depth, width, and resolution. In medical imaging, models like LB-UNet~\cite{Xu_LBUNet_MICCAI2024} reduce parameters using Group Shuffle Attention while maintaining accuracy. However, extreme parameter reduction often degrades reconstruction quality~\cite{9607618}, highlighting the trade-off between model efficiency and output fidelity, which remains a key research focus.

Hypercomplex neural networks provide a promising solution by encoding multi-channel data using algebraic structures, thereby reducing parameters while preserving representational power~\cite{grassucci2022lightweight}. Parameterized Hypercomplex Neural Networks (PHNNs)~\cite{zhang2021beyond_quaternions} generalize this approach using learnable Kronecker products, achieving expressive power with fewer parameters. Grassucci et al.~\cite{grassucci2022lightweight} extended this to convolutional layers, effectively capturing inter-channel relationships with reduced redundancy. Adopting hypercomplex representations in model networks have demonstrated superior performance in medical imaging~\cite{AHMED2025109470,sigillo2024generalizingmedicalimagerepresentations}.These successes indicate that hypercomplex networks can maintain accuracy while drastically reducing model size – a highly desirable property for accelerating MRI reconstruction.

In this paper, we propose a lightweight deep learning model for MRI reconstruction, integrating hypercomplex parameterization with efficient network design. Based on Multi-Layer Perceptron (MLP) \cite{Bourlard1994} and Window Attention \cite{liu2021swintransformerhierarchicalvision,vaswani2023attentionneed}, we introduce Kronecker-based modules: Kronecker MLP and Kronecker Window Attention, enhancing feature representation via Kronecker-Parameterized Linear Layers, and Kronecker Convolution for parameter-efficient spatial feature extraction. Embedding these modules into an end-to-end network reduces parameters and memory while maintaining high reconstruction quality.
Building on these innovations, we develop Kronecker U-Net and Kronecker SwinMR, leveraging Kronecker Convolution in a U-Net \cite{ronneberger2015unetconvolutionalnetworksbiomedical} and SwinMR \cite{huang20222Swin}, integrating Kronecker MLP and Kronecker Window Attention \cite{liu2021swintransformerhierarchicalvision} for efficient transformer-based feature extraction. These models reconstruct high-quality MRI images from undersampled data with minimal parameters, improving deployment on hardware-constrained systems and enhancing generalization by mitigating overfitting.
To the best of our knowledge, this is the first work to apply parameterized hypercomplex-based transformations, including Kronecker MLP, Kronecker Window Attention, and Kronecker Convolution, in MRI image reconstruction.

\section{Methodology}

\subsection{Overview}
Our proposed framework is built upon a set of novel Kronecker-based modules, designed to achieve parameter-efficient and high-performance MRI reconstruction. Specifically, we introduce the Kronecker MLP, Kronecker Window Attention, and Kronecker Convolution, which leverage Kronecker-Parameterized Linear Layers and Kronecker-Parameterized Convolutional Layers to enable a hypercomplex-inspired decomposition. By applying Kronecker factorization, our method significantly reduces the parameter count while maintaining expressive capacity, as illustrated in Fig.~\ref{fig:framework}. Notably, for a given hypercomplex dimension $n$, the total number of parameters is reduced to approximately $\frac{1}{n}$ of the original, making our approach particularly advantageous for deployment in hardware-constrained MRI systems. 

Building on these innovations, we develop Kronecker SwinMR and Kronecker U-Net, two MRI reconstruction models that integrate our proposed Kronecker modules. While inspired by the structures of SwinMR and U-Net, our models fundamentally differ by incorporating Kronecker-based transformations, leading to superior parameter efficiency and reconstruction quality under challenging undersampling conditions.


\subsection{Kronecker-based Parameterization}
Our Kronecker module introduces two core components: the Kronecker Linear Layer and the Kronecker Convolution Layer.

\noindent \textbf{Kronecker Linear Layer:} 
A drop-in replacement for fully-connected layers, this layer factorizes the weight matrix into a sum of Kronecker products:
\begin{equation}
H \;=\; \sum_{i=1}^{n} A[i] \otimes S[i],
\label{eq:kronecker_linear}
\end{equation}
where each \(A[i] \in \mathbb{R}^{n \times n}\) encodes the algebraic interactions and each \(S[i] \in \mathbb{R}^{\frac{\text{out}}{n}\times \frac{\text{in}}{n}}\) represents the reduced filter weights. This formulation reduces the number of parameters to approximately \(\frac{1}{n}\) of that in a conventional layer while maintaining expressive power.

In this framework, the parameterized hypercomplex multiplication (PHM)~\cite{zhang2021beyond_quaternions} layer generalizes traditional hypercomplex multiplications. When \(n = 1\), the formulation degenerates to a standard real-valued linear (or convolutional) layer, and when \(n\) takes values corresponding to well-known hypercomplex algebras (e.g., \(n=2\) for complex numbers, \(n=4\) for quaternions), the learned operations can replicate algebraic rules such as the Hamilton product. More importantly, by learning the matrices \(A[i]\) and \(S[i]\) from data, the network can adaptively determine the best algebraic interactions—even in domains where a predefined hypercomplex structure does not exist.

For example, in MRI reconstruction, our input is complex data, where the two channels (real and imaginary) can be interpreted as a complex number \(a + bi\). Setting \(n=2\) naturally couples these channels through the Kronecker-based PHM layer, ensuring that the phase information is processed jointly and effectively preserved.

\noindent \textbf{Kronecker Convolution Layer:} 
Similarly, the Kronecker Convolution Layer reconstructs the convolutional kernel via a sum of Kronecker products. Its formulation is analogous to that in Equation~\ref{eq:kronecker_linear}, with the difference that the reduced filter weights \(S[i]\) are replaced by convolutional filters \(F[i]\):
\begin{equation}
H \;=\; \sum_{i=1}^{n} A[i] \otimes F[i],
\label{eq:kronecker_conv}
\end{equation}
where each \(A[i] \in \mathbb{R}^{n \times n}\) encodes the algebraic rules and each \(F[i] \in \mathbb{R}^{\frac{\text{out}}{n}\times \frac{\text{in}}{n}\times k \times k}\) contains a fraction of the convolution filters. The convolution then proceeds as:
\begin{equation}
Y \;=\; \text{PHC}(X) \;=\; H \ast X + \text{bias},
\end{equation}
where PHC is our Parameterized Hypercomplex Convolution (PHC) layer.
\subsection{Generalization on Various Neural Network Architectures}
Furthermore, our proposed Kronecker-based parameterization can be seamlessly integrated into a wide variety of neural network architectures. Below we briefly illustrate two representative cases:

\noindent \textbf{Transformer-based Network:} 
In Transformer architectures, we replace conventional linear transformations with our PHM layers. Specifically, the multi-head self-attention mechanism \cite{vaswani2017attention} is reformulated as:
\begin{equation}
Q,\,K,\,V \;=\; \Phi\bigl(\text{PHM}(X)\bigr),
\quad
A \;=\; \text{softmax}\!\Bigl(\frac{Q\,K^\top}{\sqrt{d_k}}\Bigr)\,V,
\end{equation}
where \(\Phi(\cdot)\) denotes the appropriate splitting of the transformed input. Moreover, the MLP sub-network in the Transformer---typically two consecutive fully connected layers with a ReLU activation---is restructured as:
\begin{equation}
Y \;=\; \text{PHM}\bigl(\text{ReLU}(\text{PHM}(X))\bigr).
\end{equation}
Note that the underlying PHM operation is defined analogously to Equation~\ref{eq:kronecker_linear}.

\noindent \textbf{Convolution-based Network:} 
In convolutional architectures such as U-Net, standard 2D convolutions
\[
Y \;=\; \text{Conv2D}(X,\,W,\,\text{bias})
\]
are replaced by our PHC layer. Here, the convolutional kernel is reconstructed as in Equation~\ref{eq:kronecker_conv}.
This substitution not only reduces the parameter count by roughly a factor of \(n\) but also enhances the ability to capture inter-channel correlations.

Overall, our work pioneers the integration of Kronecker-parameterized hypercomplex layers into both Transformer-based and convolution-based architectures for MRI reconstruction. This contribution establishes a new paradigm for efficient deep learning models in medical imaging, demonstrating the versatility and impact of our proposed approach.

\subsection{Model Optimization}
We employ a composite loss function that integrates image-domain consistency, frequency-domain fidelity and perceptual quality to ensure high-quality MRI reconstruction. Specifically, for a given ground-truth MRI image $x$ and its reconstruction $\hat{x}$, the total loss consists of Charbonnier loss applied in both the image domain and frequency domain to enforce structural and spectral consistency, and $\ell_1$ loss in the latent space, computed via a pretrained VGG~\cite{Simonyan2014VeryDC} feature extractor, to preserve perceptual quality.  
This composite loss formulation ensures robust and high-fidelity reconstruction, even under challenging undersampling conditions.

\section{Experimental Setting}

\noindent \textbf{Dataset:} Our evaluation framework employed the FastMRI open repository \cite{zbontar2019fastmriopendatasetbenchmarks}, containing single-coil complex-valued MRI acquisitions. From the original training/validation cohorts, we curated 684 non-fat-suppressed proton-density weighted knee MRI studies, implementing stratified partitioning with approximate 6:1:3 ratios - 420 for training, 64 for validation, and 200 holdout test cases. Each volumetric scan underwent standardized preprocessing: selection of 20 central coronal-plane slices in complex-valued format followed by spatial normalization through \(320\times320\) center cropping.
\begin{figure}[htbp]
    \centering
    \includegraphics[width=0.85\textwidth]{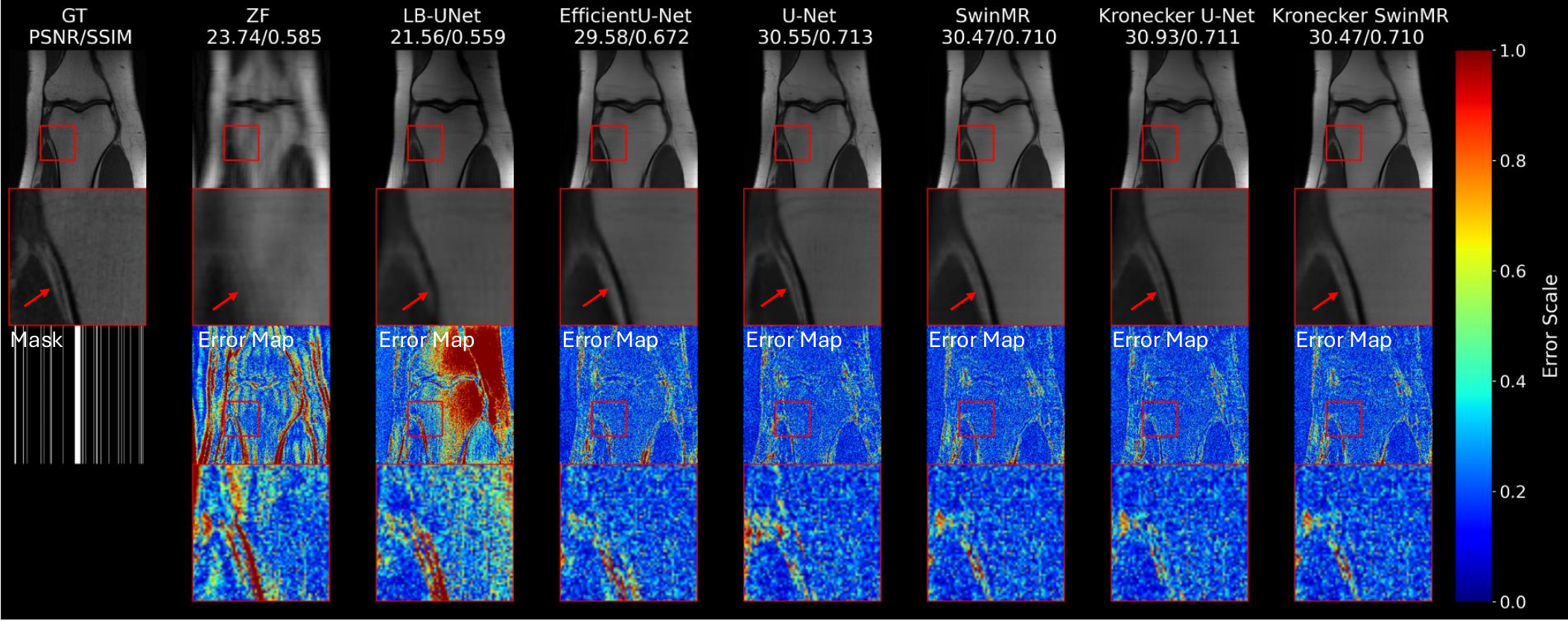}
    \caption{Comparison of reconstruction results (acceleration factor=16$\times$). Ground truth (GT) and zero-filled (ZF) reconstruction vs. LB-UNet, EfficientU-Net~\cite{tan2019efficientnet}, U-Net, SwinMR, our proposed Kronecker U-Net and Kronecker SwinMR.}
    \label{fig:comparison}
\end{figure}
The single-coil emulated data provided in FastMRI was used as the complex-valued ground. Undersampling patterns were synthesized using the FastMRI reference implementation \cite{zbontar2019fastmriopendatasetbenchmarks}, with Cartesian sampling \cite{feng2022golden} schemes at acceleration factors \( R \in \{8, 16\} \) uniformly applied throughout all experimental protocols.


\noindent \textbf{Implementation Details:} We set number of the Kronecker Layer to 2 both in Kronecker SwinMR and Kronecker UNet since the input channels are the real and imaginary parts of the complex. The loss function incorporates weighting parameters $\alpha$, $\beta$, $\gamma$, and $\eta$, which are set to 15, 0.1, 0.0025, and 0.1, respectively.
\begin{table}[!t]
    \centering
    \caption{Acceleration factor=8$\times$: Performance Comparison on FastMRI dataset}
    \label{tab:combined_perf_comparison_af8}
    \small 
    \renewcommand{\arraystretch}{0.8} 
    \setlength{\tabcolsep}{4pt} 
    \resizebox{\textwidth}{!}{%
    \begin{tabular}{lrrrrrr}
    \toprule
    \textbf{Model} & 
    \textbf{Params (M) $\downarrow$} & 
    \textbf{SSIM (Std.) $\uparrow$} & 
    \textbf{PSNR (Std.) $\uparrow$} & 
    \textbf{LPIPS (Std.) $\downarrow$} \\
    \midrule
    Zero-Filled           & -     & 0.430 (0.098) & 22.75 (1.73) & 0.504 (0.058) \\
    \midrule
    LB UNet               & 1.025 & 0.430 (0.130) & 23.92 (2.54) & 0.449 (0.067) \\
    Efficient U-Net       & 3.896 & 0.730 (0.083) & 28.97 (2.41) & 0.396 (0.066) \\
    \midrule
    U-Net                 & 5.654 & 0.783 (0.067) & 29.64 (2.47) & 0.290 (0.069) \\
    Kronecker U-Net       & 2.630 & 0.762 (0.075) & 29.78 (2.52) & 0.274 (0.006) \\
    \midrule
    SwinMR                  & 2.380 & 0.771 (0.073) & 29.65 (2.41) & 0.265 (0.050) \\
    Kronecker SwinMR        & 1.204 & 0.763 (0.073) & 29.35 (2.35) & 0.266 (0.048) \\
    \bottomrule
    \end{tabular}%
    }
\end{table}
All experiments are performed on four NVIDIA GeForce RTX 4090 with 24GB GPU memory each and evaluated on a single NVIDIA RTX 4090 GPU. All models underwent training for 100,000 gradient steps, utilizing the Adamoptimiser~\cite{diederik2014adam} with a learning rate of 2$\times$$10^{5}$ and a batch size of 8. For quantitative analysis, we employed Peak Signal-to-Noise Ratio (PSNR), Structural Similarity Index Measure (SSIM)~\cite{wang2004ssim}, and Learned Perceptual Image Patch Similarity (LPIPS)~\cite{Zhang2018Unreasonable_LPIPS} to assess reconstruction quality. 



%
%

\section{Result and Discussion}

\subsection{Comparison with other methods}

Table~\ref{tab:combined_perf_comparison_af8} presents the performance comparison on the FastMRI dataset with an acceleration factor of $8\times$. The Kronecker Parameterized Layers introduced a marginal degradation in PSNR, SSIM and LPIPS for both U-Net and SwinMR, but this difference was not statistically significant ($p > 0.05$), indicating no compromise in performance. In contrast, compact models from previous literature (LB-UNet and Efficient U-Net) showed significantly degraded PSNR, SSIM and LPIPS values ($p < 0.05$). \textbf{Notably, without compromising performance results, Kronecker variants achieved a dramatic reduction in model parameters by 2-fold.}

\begin{table}[htbp]
    \centering
    \caption{Acceleration factor=16$\times$: Performance Comparison on FastMRI dataset}
    \label{tab:perf_comparison_af16}
    \small 
    \renewcommand{\arraystretch}{0.8} 
    \setlength{\tabcolsep}{4pt} 
    \resizebox{\textwidth}{!}{%
    \begin{tabular}{lrrrrrr}
    \toprule
    \textbf{Model} & 
    \textbf{Params (M) $\downarrow$} & 
    \textbf{SSIM (Std.) $\uparrow$} & 
    \textbf{PSNR (Std.) $\uparrow$} & 
    \textbf{LPIPS (Std.) $\downarrow$} \\
    \midrule
    Zero-Filled             & -     & 0.544 (0.058) & 22.61 (1.73) & 0.580 (0.049) \\
    \midrule
    U-Net                   & 5.654 & 0.735 (0.064) & 28.12 (2.14) & 0.347 (0.007) \\
    Kronecker\ UNet         & 2.630 & 0.707 (0.071) & 27.89 (2.15) & 0.339 (0.006) \\
    \midrule
    SwinMR                  & 2.380 & 0.707 (0.070) & 27.47 (2.06) & 0.332 (0.052) \\
    Kronecker\ SwinMR       & 1.204 & 0.680 (0.068) & 26.59 (1.96) & 0.353 (0.050) \\

    \bottomrule
    \end{tabular}%
    }
\end{table}

A further evaluation was conducted on U-Net~\cite{ronneberger2015unet}, SwinMR~\cite{huang2022swintransformerfastmri}, and their respective Kronecker variants at a higher acceleration factor of $16\times$. As shown in Table~\ref{tab:perf_comparison_af16}, where same conclusions were reproduced: no statistically significant difference observed in performance metrics between the Kronecker variants and the original models (SwinMR and U-Net), with the Kronecker variants achieving a substantial reduction in model parameters. 

\subsection{Ablation for Higher Hypercomplex Dimensions}

As discussed in Section~2.1, the hypercomplex dimension ($n$) is inversely related to the model parameter count, with higher values of $n$ leading to greater parameter reduction. To investigate the impact of this further, additional experiments were conducted on the Kronecker U-Net variant to evaluate whether increasing $n$ would compromise performance. As presented in Table~\ref{tab:perf_comparison_ablation}, setting $n=4$ slightly degraded PSNR, SSIM and LPIPS values. However, statistical analysis using the t-test indicated that these differences were not significant.  This suggests that increasing the hypercomplex dimension can achieve substantially greater parameter reduction but without significantly affecting model performance. Consequently, this approach provides a promising avenue for developing even more compact models by exploring higher hypercomplex dimensions (e.g., $n=8$ or $n=16$), potentially leading to more memory efficient architectures.

\begin{figure}[t!]
    \centering
    \includegraphics[width=0.8\textwidth]{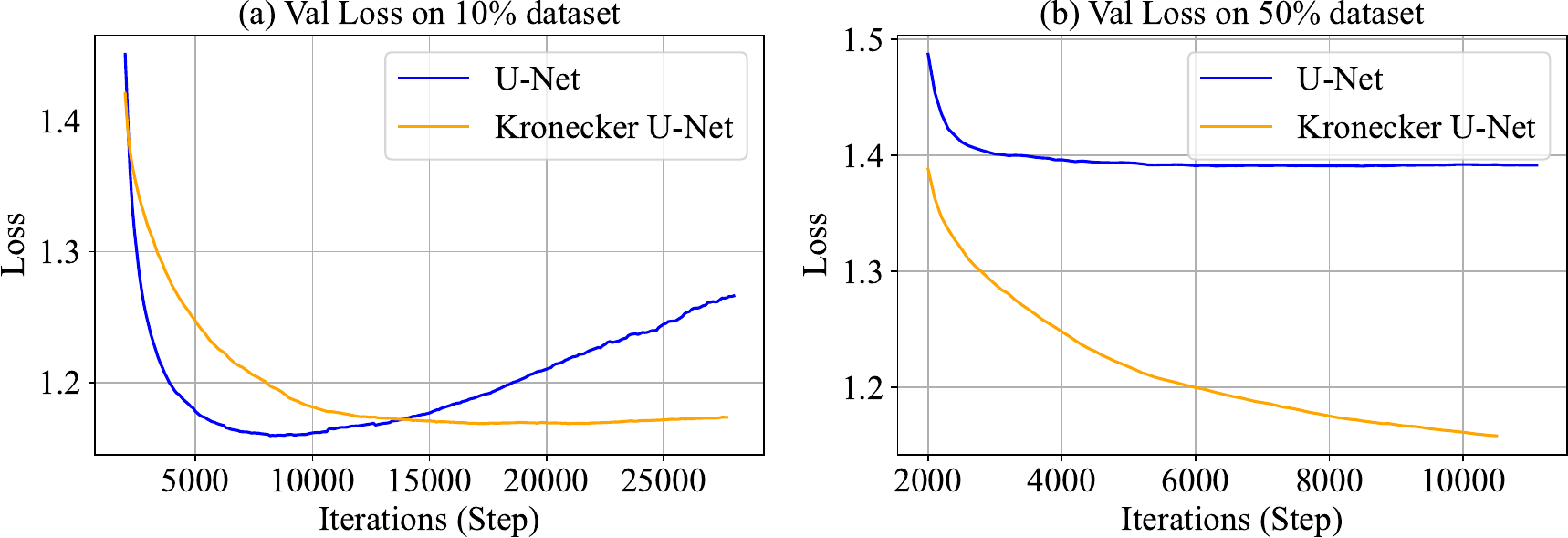}
    \caption{Validation loss curves for U-Net and Kronecker U-Net trained on (a) 10\% and (b) 50\% of the data respectively. x-axis: \# iterations; y-axis: validation loss.}
    \label{fig:val_loss_overfitting}
\end{figure}

\subsection{Generalization over Limited Dataset}

\begin{table}[htbp]
    \centering
    \caption{Ablation Study with Different Hypercomplex Dimensions (n), AF=8$\times$}
    \label{tab:perf_comparison_ablation}
    \small 
    \renewcommand{\arraystretch}{0.8} 
    \setlength{\tabcolsep}{4pt} 
    \resizebox{\textwidth}{!}{%
    \begin{tabular}{lrrrrrr}
    \toprule
    \textbf{Model} & 
    \textbf{Params (M) $\downarrow$} & 
    \textbf{SSIM (Std.) $\uparrow$} & 
    \textbf{PSNR (Std.) $\uparrow$} & 
    \textbf{LPIPS (Std.) $\downarrow$} \\
    \midrule
    Zero-Filled           & -     & 0.430 (0.098) & 22.75 (1.73) & 0.504 (0.058) \\
    SwinMR                      & 2.380 & 0.771 (0.073) & 29.65 (2.41) & 0.265 (0.050) \\
    Kronecker\ SwinMR n=2       & 1.204 & 0.763 (0.073) & 29.35 (2.35) & 0.266 (0.048) \\
    Kronecker\ SwinMR n=4       & 0.746 & 0.757 (0.072) & 29.02 (2.26) & 0.273 (0.047) \\
    \bottomrule
    \end{tabular}%
    }
\end{table}

A common limitation of parameter-heavy models is their tendency to overfit on smaller datasets due to their complexity. In contrast, memory-efficient models, such as the proposed Kronecker variants, enhance generalization with significantly fewer parameters. To evaluate this, experiments were conducted using reduced training subsets, tracking validation loss curves to assess overfitting behavior. As shown in Fig.~\ref{fig:val_loss_overfitting}, the Kronecker variant consistently achieved lower validation loss compared to the original U-Net, where the U-Net plateaued and overfit, especially when trained on just 10\% of the data. In contrast, the Kronecker variant maintained stable convergence without overfitting. These findings demonstrate the superior generalization capability of Kronecker variants, particularly when trained on limited data, making them a robust choice for scenarios where large-scale datasets are unavailable. This resilience against overfitting is especially beneficial for state-of-the-art models prone to over-parameterization.

\section{Conclusion}

This study introduced Kronecker-Parameterized Hypercomplex Layers for MRI reconstruction, achieving roughly 50\% fewer parameters while maintaining high reconstruction quality. By integrating Kronecker MLP, Kronecker Window Attention, and Kronecker Convolution into U-Net and SwinMR, our models demonstrated competitive PSNR, SSIM, and LPIPS metrics. Experiments showed no significant performance drop at high acceleration factors ($8\times$ and $16\times$). Increasing the hypercomplex dimension ($n$) further reduced parameters without compromising performance. Additionally, Kronecker variants exhibited better generalization and reduced overfitting on limited datasets. This approach enables efficient MRI reconstruction on resource-constrained hardware.

\section*{Acknowledgments}
Yinzhe Wu was supported in part by the Imperial College London I-X Moonshot
Seed Fund and in part by the Imperial College London President’s PhD Scholarship.
Guang Yang was supported in part by the ERC IMI (101005122), the H2020
(952172), the MRC (MC/PC/21013), the Royal Society (IEC NSFC 211235), the
NVIDIA Academic Hardware Grant Program, the SABER project supported by
Boehringer Ingelheim Ltd, NIHR Imperial Biomedical Research Centre (RDA01),
Wellcome Leap Dynamic Resilience, UKRI guarantee funding for Horizon Europe
MSCA Postdoctoral Fellowships (EP/Z002206/1), UKRI MRC Research Grant,
TFS Research Grants (MR/U506710/1), and the UKRI Future Leaders Fellowship
(MR/V023799/1).

    



%
%
\bibliographystyle{splncs04}
\bibliography{bibliography}

\end{document}